\documentstyle[twocolumn,prl,aps,epsf]{revtex}
\begin{document}

\twocolumn[\hsize\textwidth\columnwidth\hsize\csname
@twocolumnfalse\endcsname

\title{Ultrafast Spin 
Dynamics in GaAs/GaSb/InAs Heterostructures
Probed by Second Harmonic Generation} 
\author{Yu. D. Glinka, T. V. Shahbazyan, I. E. Perakis and N. H. Tolk}
\address{Department of Physics and Astronomy, Vanderbilt University,
  Nashville, TN 37235}   
\author{X. Liu, Y. Sasaki and J. K. Furdyna}
\address{Department of Physics, University of Notre Dame, Notre Dame,
IN 46556}
\date{\today}
\maketitle
\draft
\begin{abstract}
We report the first application of pump-probe second harmonic generation
(SHG) measurements to characterize optically-induced magnetization in
non-magnetic multilayer semiconductors. In the experiment, coherent spins are
selectively excited by a pump beam in the GaAs layer of GaAs/GaSb/InAs
structures.  However, the resulting net magnetization manifests itself through
the induced SHG probe signal from the GaSb/InAs interface, thus indicating a
coherent spin transport across the heterostructure.  We find that the
magnetization dynamics is governed by an interplay between the spin density
evolution at the interfaces and the spin dephasing.
\end{abstract}
\pacs{PACS numbers:}

\vskip2pc]

Ultrafast spin-sensitive spectroscopy provides unique information
about spin decoherence in semiconductor heterostructures
as well as spin-polarized transport across interfaces. Knowledge of
the processes governing spin dynamics is essential for designing novel
multifunctional electronic and optoelectronic devices, including base
components for quantum computing\cite{awschalom99,awschalom01}. 
Among the wide variety of multilayer semiconductor systems, GaSb/InAs
heterostructures are especially promising because of their unusual band
alignment\cite{koshihara97}.

The excitation of an ensemble of coherent spins by a circularly
polarized laser light at the photon energy just above the bandgap gives rise
to a net magnetization.
The techniques typically used to monitor the spin dynamics, such as
polarized photoluminescence 
spectroscopy\cite{awschalom99,smith93,crooker96,heberle94}, pump-probe
transmission/reflection\cite{tackeuchi97,ohno99,hall99,boggess00}, and
time-resolved Faraday or Kerr rotation\cite{awschalom99,awschalom01},
all rely on the {\em linear} response of the spin subsystem to the
probing light. On the other hand, the {\em nonlinear} optical effects,
such as second harmonic generation (SHG), are known to be highly
sensitive to local magnetic fields occurring at magnetized surfaces
and at interfaces in magnetic-semiconductor-based
multilayers\cite{liebsch88,pan89,reif91,wu00,gridnev01}.

In this Letter we report the first application of ultrafast pump-probe SHG
measurements to characterize optically-induced magnetization in non-magnetic
GaAs/GaSb and GaAs/GaSb/InAs heterostructures. A normally-incident circularly
polarized pump beam of 150-fs pulse duration is used to selectively create an
excess 
of spin-polarized electrons in the GaAs epilayer.  A linearly polarized probe
beam from the same source is used to measure SHG signal in reflection
geometry. The pump-induced SHG signal, which is monitored as a function of
probe-to-pump delay times, is due to interfacial electric and magnetic fields
created by the pump-excited carriers. Only the GaAs/GaSb/InAs samples show a
significant induced magnetization, indicating a {\em coherent spin transport}
across the heterostructure.  The temperature dependence of the induced SHG
signals in the range from 4.3 to 300 K reveals two distinct
processes that affect the magnetization dynamics: the evolution of the local
spin density at the interfaces and the spin dephasing.

We have investigated four heterostructures grown by molecular beam epitaxy:
(1) GaAs/GaSb(400 nm); (2) GaAs/GaSb(20 nm); (3) GaAs/GaSb(500 nm)/InAs(20
nm), with an InSb interface between GaSb and InAs layers; and (4)
GaAs/GaSb(500 nm)/InAs(20 nm), with a GaAs interface between GaSb and InAs.
All samples were grown on semi-insulating (100) GaAs substrates.  Prior to
GaSb or GaSb/InAs deposition the substrates were cleaned {\em in situ} by
oxide desoprtion, by heating to 600 $^{\circ}$C, after which a 100 nm GaAs
buffer layer was grown at 590 $^{\circ}$C. In deposition of the overlayers, we
used 490 $^{\circ}$C for GaSb growth and 450 $^{\circ}$C for InAs. All
optical measurements were carried out in a liquid helium cryostat.  The
initial beam of 150-fs pulses from a mode-locked Ti:Al$_2$O$_3$ laser (Mira
900) at the wavelength of 800 nm (1.55 eV) and a repetition rate of 76 MHz was
split into pump and probe beams.  The probe beam of 120-mW-average power has
passed through an optical delay stage. The pump beam chopped at a frequency of
400 Hz was of the same average power. The overlap spot of the pump and probe
beams on the sample was $\sim 100$ $\mu$m in diameter.  The pump beam was
incident normally on the sample with either left- or right-handed circular
polarization ($\sigma^{-}$ or $\sigma^{+}$, respectively). The probe beam was
linearly-polarized ({\em p} or {\em s}), and was directed to the sample
surface at the angle of 85$^{\circ}$.  The SHG signal was optically separated
from the reflected fundamental probe beam and measured by a photomultiplier
tube through a ``lock-in'' amplifier triggered by the chopped pump pulses.

Figure 1 shows the pump-induced SHG signals taken on the GaAs/GaSb
heterostructure (Sample 1) at a temperature $T= 4.3$ K.  No significant
difference was observed between signals measured with $\sigma^{-}$ or
$\sigma^{+}$-polarized pump light [Figs.\ 1(a)\ and\ 1(b), respectively]. Note
that only the {\em p}-linearly-polarized (in the plane of incidence) probe
light contributes to the induced signal, in agreement with SHG measurements on
magnetized surfaces\cite{reif91}. The measured signal was fitted by a combined
exponential rise/decay function, as shown in Figs.\ 1(a)\ and\ 1(b). The signal
intensity increases with a time constant of $\tau_{R1} \sim 3$ ps, followed by
a decay with $\tau_D \sim 100$ ps. The induced signal completely disappears at
room temperature [Fig. 1(a)]. The temperature dependence of the peak intensity
in the range from 4.3 to 300 K reveals two drops, at $\sim 40$ and $\sim 170$
K [Fig.\ 1(c)]. The rise-time constant $\tau_{R1}$ decreases with temperature,
reaching $\sim 300$ fs above 200 K. In contrast, the decay-time $\tau_D$ is
almost unchanged over the entire range [Fig.\ 1(d)].  The signal becomes
completely undetectable when the thickness of GaSb layer is reduced to 20 nm
(Sample 2), indicating that GaSb is essential for the process in question.

The behavior of the induced SHG signal from GaAs/GaSb/InAs samples is more
complex.  The signal appears to be stronger in the heterostructure with an
InSb interface, as compared to that with a GaAs interface 
[Figs.\ 2(a)\ and\ 3(a)].  Additionally, now a long-lived $\tau_{R2}\sim 15$
ps rise-time component arises, resulting in a shift of the signal peak towards
longer times with respect to those for the GaAs/GaSb samples.  Moreover, the
SHG signal intensities for $\sigma^{-}$ and $\sigma^{+}$ pump polarizations
are now different, indicating the presence of an induced magnetization.  A 
striking feature is a rapid increase of the peak intensity with temperature in
the range from 4.3 to 50 K, followed by a plateau and a subsequent decrease
above  100 K [Figs.\ 2(c)\ and\ 3(c)].  The temperature behavior of the
initial rise-time constant closely matches that of $\tau_{R1}$ in the
GaAs/GaSb samples. However, the decay-time constant noticeably decreases for
$T \gtrsim 180$ K, reaching $\sim 20$ ps at room temperature.

Comparison of SHG signals taken for GaAs/GaSb sample with different pump
polarization shows no marked induced magnetization (Fig.\ 1), so the induced
signal is due to the {\em electric} field at the interfaces. The interfacial
electric fields caused by a charge separation among photoexcited carriers are
known to strongly enhance the SHG response\cite{lupke99}.  In sharp contrast,
the measured SHG signal for GaAs/GaSb/InAs samples shows a significant 
{\em net magnetization}, which we ascribe to the presence of coherent spins in
the InAs layer. Since the spins were excited in the GaAs layer, 
this indicates an {\em interlayer coherent spin transport}.
Correspondingly, the long-lived $\tau_{R2}\sim 15$ ps
rise-time component characterizes the rate of spin transfer to the InAs layer,
which slightly increases with temperature [Figs.\ 2(d)\ and\ 3(d)].

Retaining only linear terms in the induced electric field,
${\cal E}(t)$, and magnetic field, $M(t)$, the nonlinear pump-probe
polarization can be presented as\cite{bennemann99,lupke99}
\begin{equation}
\label{pol}
P_{\pm}^{NL}(2\omega,t) =\Bigl[\chi^{(2)}+\chi_{e}^{(3)}{\cal E}(t)
\pm\chi_{m}^{(3)}M(t)\Bigr][E(\omega)]^2,
\end{equation}
where $E(\omega)$ is the electric field component of the incident probe light, 
and $\chi^{(2)}$, $\chi_{e}^{(3)}$, and $\chi_{m}^{(3)}$ are the
corresponding nonlinear susceptibilities.  The alternate signs in 
Eq.\ (\ref{pol}) correspond to the two possible directions of the induced
magnetic field normal to the interface.  The magnetic- and
electric-field-induced contributions are then extracted from the pump-induced
SHG signal intensity, $\Delta I_{\pm}^{(2\omega)}(t)$, as
\begin{eqnarray}
\label{el-mag}
\Delta I_{-}^{(2\omega)}-\Delta I_{+}^{(2\omega)}\propto M(t),
 ~~~ 
\Delta I_{-}^{(2\omega)}+\Delta I_{+}^{(2\omega)} \propto {\cal E}(t).
\end{eqnarray}
Remarkably, the extracted induced magnetization curves [Figs.\ 2(b)\ and\ 3(b)]
are similar for samples with either type of interface between GaSb
and InAs layers, although total intensities differ considerably.  The
profile of the extracted electric-field-induced signal closely follows that
measured for the GaAs/GaSb samples, but the amplitude is different.  Thus the
electric- and magnetic-field-induced contributions appear to be additive,
indicating that the higher-order nonlinear terms, neglected in 
Eq.\ (\ref{el-mag}), are indeed small.

Because the laser light was tuned just above the GaAs bandgap, spin-polarized
electrons excited in the smaller bandgap GaSb and InAs layers are much more
energetic (0.74 and 1.11 eV, respectively). These electrons lose their spin
coherence on the time scale of the pump pulse duration due to the strong $E^3$
energy dependence of the Dyakonov-Perel (DP) spin-orbit scattering
process\cite{book}.  Due to the band alignment of GaAs/GaSb/InAs
heterostructure, the incoherent electrons accumulate in the GaAs and InAs
regions while the holes, which lose coherence much faster than the
electrons\cite{tackeuchi97,ohno99,hall99,boggess00}, are amassed in the GaSb
layer. This leads to a charge separation at the interfaces (Fig.\ 4). The rise
of the interfacial electric fields manifests itself in the initial growth of
the SHG signal ($\sim 3$ ps at 4.3 K).  The rise-time decreases to 
$\sim 300$ fs for $T> 200$ K, matching the
typical room-temperature values\cite{tackeuchi97,ohno99,hall99,boggess00}.
The induced electric fields at the interfaces bend the initial energy profile,
lowering the barrier at the GaAs/GaSb interface (Fig.\ 4). As negative charges
overcome the barrier and transfer to the InAs layer, the electric field at the
GaAs/GaSb interface decreases, while it increases at the GaSb/InAs interface.
This redistribution of negative charges occuring on the time scale of 
$\sim 15$ ps is similar to that of coherent spins.  A subsequent 
relaxation of the interfacial electric fields manifests itself as the 
$\tau_D \sim 100$ ps decay of the induced SHG signal\cite{glinka}.
The residual electric field at the interfaces contributes to the signal as a
long-time constant background observed for GaAs/GaSb/InAs samples.  Note that
in the GaAs(100 nm)/GaSb(20 nm) sample the carriers are accumulated
predominantly in the GaAs layer, so the interfacial fields are insignificant
and the corresponding SHG signal becomes undetectable.

The fraction of the spin-polarized electrons in the GaAs layer was about 50\%,
since the relative concentration of excited heavy and light holes is estimated
as 3:1. The fraction reduces to less than 8\% when incoherent electrons from
other layers are taken into account.  Due to the smallness of this fraction,
the magnetization in GaAs/GaSb samples is weak. However, the GaAs/GaSb/InAs
samples show a relatively strong magnetization. This results from the large
electric field gradient confining the spins at the GaSb/InAs interface.

The measured temperature dependence of the SHG signal supports the above
interpretation. For both GaAs/GaSb and GaAs/GaSb/InAs samples, the amplitudes
of the induced signal exhibit a rather remarkable behavior in the range from
4.3 to 100 K. For GaAs/GaSb samples, the peak intensity drops sharply to a
certain level, while it grows and then stabilizes for GaAs/GaSb/InAs samples
[Figs.\ 1(c),\ 2(c),\ and\ 3(c)]. We attribute these features to the ability of
thermally activated electrons in GaAs to overcome the GaAs/GaSb interface
barrier.  Since the efficiency of this process increases with temperature, the
signal amplitude in GaAs/GaSb samples experiences a drop as the interfacial
electric field weakens.  Correspondingly, in GaAs/GaSb/InAs samples, the
magnetic field at the GaSb/InAs interface increases with temperature because
of the arrival of additional spins. Further decrease in the induced SHG
intensity in the temperature range from 100 to 300 K indicates a weakening of
the interfacial fields as the electron wave-functions become more extended,
effectively reducing the charge density at the interface.

Since electrons with uncompensated spin in GaAs occupy higher-lying energy
levels, the effective barrier for them is lower than that for incoherent 
electrons. The activation of spin-polarized electrons leads to the initial
rise of the signal peak intensity with temperature in GaAs/GaSb/InAs samples
[Figs.\ 2(c)\ and\ 3(c)]. The subsequent signal stabilization in the
range from 50 to 100 K is due to the competing process involving a decrease in
the interfacial electric field as the incoherent electrons begin to pass
through the barrier. Correspondingly, the induced magnetization grows
near-linearly with temperature in the range from 4.3 to 100 K. Using a simple
model, which takes into account the thermal activation of carriers at the
interface, the effective barrier height values were estimated as $\sim 2$ meV
and $\sim 8$ meV for spin-polarized and incoherent electrons, respectively.
Note that the induced signal in GaAs/GaSb samples also shows a plateau in the
range from 100 to 170 K.  This can be attributed to the thermal
activation of electrons previously trapped at the impurity centers in bulk
GaAs.

The temperature dependence of decay-time constants offers a clearer view as of
which processes contribute to the magnetization dynamics.  For all the
samples, the decay times for both electric- and magnetic-field-induced signals
are stable (at $\tau_D\sim 100$ ps) in the temperature range
from 4.3 to 180 K [Figs.\ 1(d),\ 2(d),\ and 3(d)].  At higher temperatures,
however, $\tau_D$ sharply decreases for the GaAs/GaSb/InAs samples, while
it remains unchanged for the GaAs/GaSb samples. The dominant source for
spin-flip relaxation at $T\gtrsim 100$ K is known to be the DP mechanism,
which yields $T^3$ dependence of the spin-orbit scattering rate\cite{book}.
Assuming the above dependence, the measured 20 ps decay-time constant  at room
temperature scales to $\sim 100$ ps when the temperature is reduced to 180 K, 
indicating that the DP mechanism is prevalent in the range from 180 to 300 K.
Below 180 K, a decrease in the spin density at the GaSb/InAs interface due to 
the decay of interfacial electric fields dominates over the spin dephasing, 
being a primary source of the magnetization dynamics.

In summary, pump-probe SHG measurements for non-magnetic GaAs/GaSb/InAs
semiconductor heterostructures revealed interlayer coherent spin transport.
We found that the optically-induced magnetization dynamics in such structures
originates from two distinct sources: one of them related to the evolution of
the local spin density at the interfaces, and the other one arising from the
spin dephasing.  The extreme sensitivity of the SHG to the interfacial fields,
which allowed us to distinguish between these two mechanisms, makes it a
unique tool for studying the spin and carrier dynamics in multilayer
semiconductors.

This work was supported by ONR and by the DARPA/SPINS Program.


\begin{figure}
\caption{Pump-induced SHG signal from GaAs/GaSb 
sample measured at 4.3 K (in green) and 
300 K (in blue) for   $\sigma^{-}$-polarized (a) and 
$\sigma^{+}$-polarized (b) pump.  
The fit with exponential rise/decay functions is shown by red solid curves. 
Inset (c): temperature dependence of the maximum signal 
intensity ($\Delta I_{max}$) and its fit (solid line), as explained in the 
text.
Inset (d): temperature dependence of rise-time ($\tau_{R1}$) and decay-time 
($\tau_D$) constants.  
}
\label{fig:1}
\end{figure}

\begin{figure}
\caption{Pump-induced SHG signal from
GaAs/GaSb/InAs sample with an InSb interface between GaSb and InAs, measured
at 4.3 K with 
$\sigma^{+}$ and $\sigma^{-}$ pump polarization (a)
and the extracted magnetization (b), obtained as difference of
signals in (a).
Inset (c): temperature dependence of the maximum signal 
intensity ($\Delta I_{max}$) (in red) and extracted magnetization (in blue).
The fit to data is shown by solid line.
Inset (d): temperature dependence of rise-time ($\tau_{R1}$ and
$\tau_{R1}$) and decay-time ($\tau_D$) constants for signal taken with
$\sigma^{-}$-polarized pump in (a).
}
\label{fig:2}
\end{figure}
\begin{figure}
\caption{Same as Fig.\ \ref{fig:2} but for GaAs/GaSb/InAs sample with a GaAs
interface between GaSb and InAs.
}
\label{fig:3}
\end{figure}

\begin{figure}
\caption{The initial bandgap energy alignment for GaAs/GaSb/InAs
heterostructure and its realignment due to induced electric
fields at the interfaces.
}
\label{fig:4}
\end{figure}

\begin{references}

\bibitem{awschalom99}D. D. Awschalom and J. M. Kikkawa, 
Physics Today {\bf 33} (June 1999).

\bibitem{awschalom01}I. Malajovich {\em et al.}, 
Nature {\bf 411}, 770 (2001).

\bibitem{koshihara97}S. Koshihara {\em et al.},
Munekata,
Phys. Rev. Lett. {\bf 78}, 4617 (1997).


\bibitem{smith93}J. F. Smyth {\em et al.},
Phys. Rev. Lett. {\bf 71}, 601 (1993).

\bibitem{crooker96}S. A. Crooker {\em et al.},
Phys. Rev. Lett. {\bf 77}, 2814 (1996).


\bibitem{heberle94}A. P. Heberle, W. W. Ruhle, and K. Ploog, 
Phys. Rev. Lett. {\bf 72}, 3887 (1994).

\bibitem{tackeuchi97}A. Tackeuchi, O. Wada, and Y. Nishikawa, 
Appl. Phys. Lett. {\bf 70}, 1131 (1997).

\bibitem{ohno99}Y. Ohno {\em et al.}, 
Phys. Rev. Lett. {\bf 83}, 4196 (1999).

\bibitem{hall99}K. C. Hall {\em et al.}, 
Appl. Phys. Lett. {\bf 75}, 3665 (1999).


\bibitem{boggess00}T. F. Boggess {\em et al.}, 
Appl. Phys. Lett. {\bf 77}, 1333 (2000).

\bibitem{liebsch88}A. Liebsch, Phys. Rev. Lett. {\bf 61}, 1233 (1988).

\bibitem{pan89}R. P. Pan, H. D. Wei, and Y. R. Shen, 
Phys. Rev. B {\bf 39}, 1229 (1989).

\bibitem{reif91}J. Reif {\em et al.}, 
Phys. Rev. Lett. {\bf 67}, 2878 (1991).

\bibitem{wu00}Y. Z. Wu {\em et al.}, 
Phys. Rev. B {\bf 63}, 054401 (2000).

\bibitem{gridnev01}V. N. Gridnev {\em et al.}, 
Phys. Rev. B {\bf 63}, 184407 (2001).
 
\bibitem{lupke99}G. Lupke, Surf. Sci. Rep. {\bf 35}, 75 (1999).

\bibitem{bennemann99}K. H. Bennemann,
J. Magn. Magn. Mater. {\bf 200}, 679 (1999).

\bibitem{book}M. I. Dyakonov and V. I. Perel, in {\em Optical Orientation},
edited by F. Meyer and B. P. Zakharchenya (North Holland, Amsterdam 1984),
p. 11; G. E. Pikus and A. N. Titkov, {\em ibid.} p. 73.

\bibitem{glinka} The dynamics of the interfacial electric fields was
studied using linearly polarized pump beam and will be published elsewhere.
\end{references}
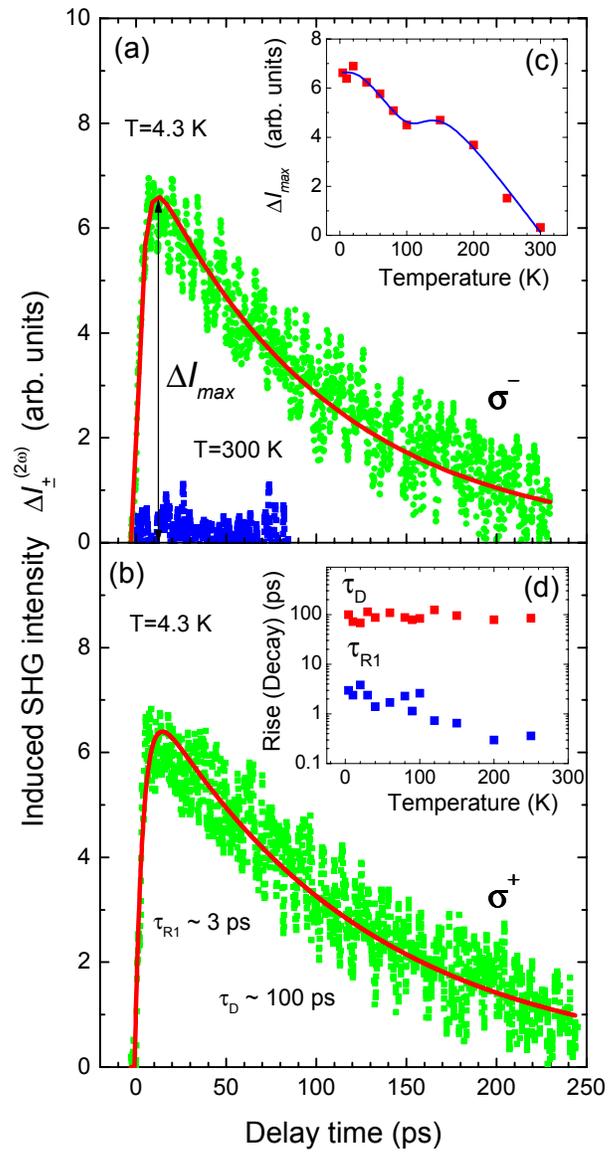
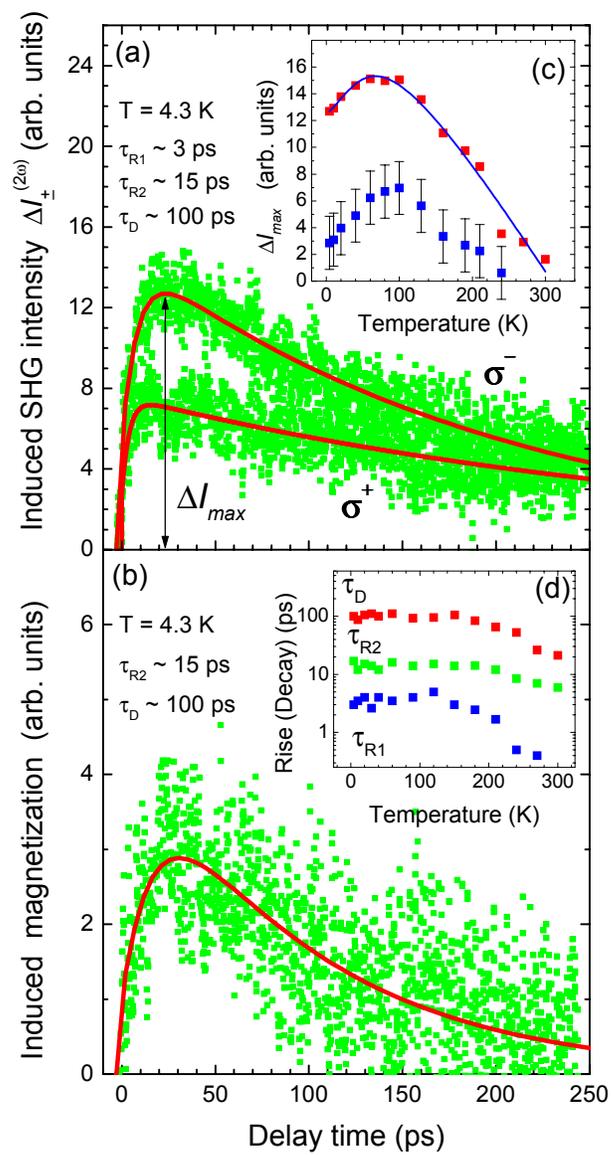
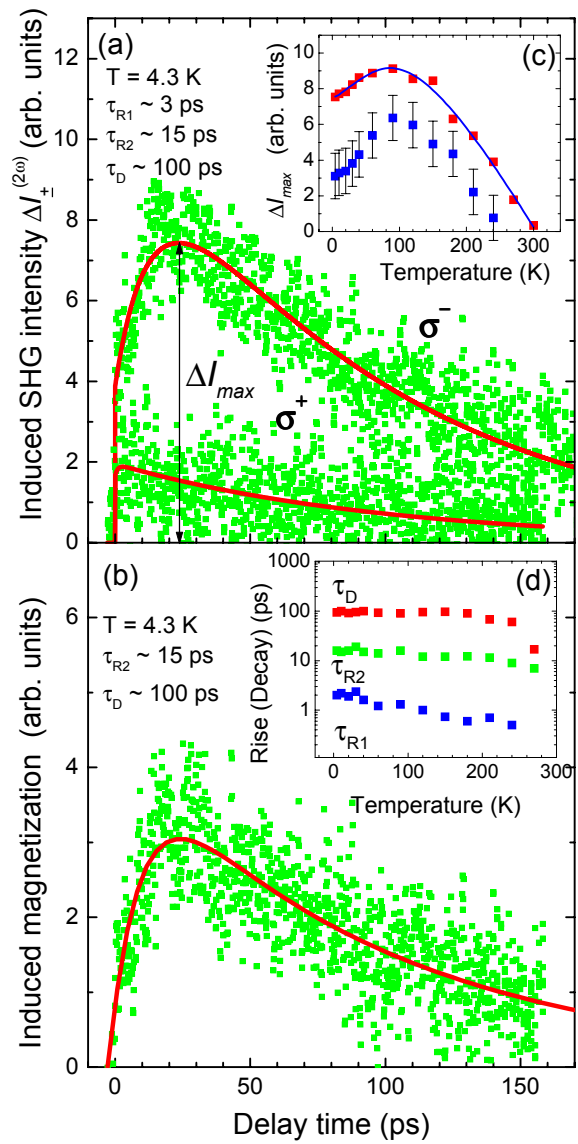
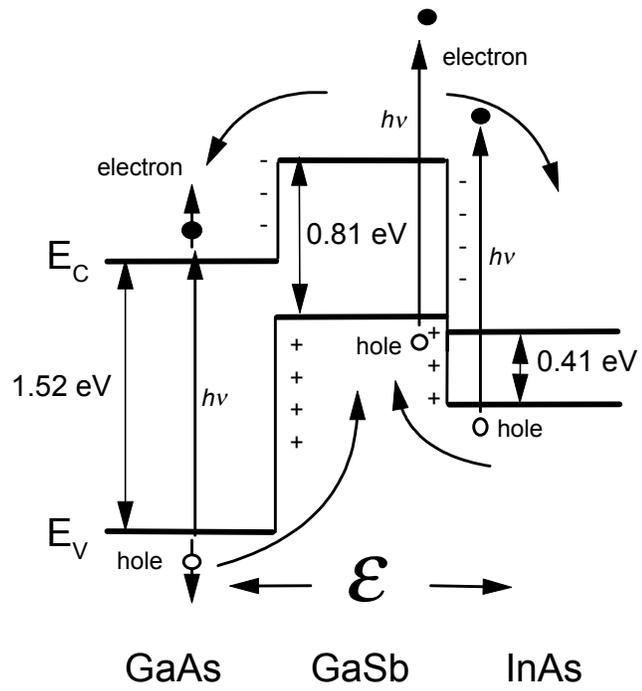
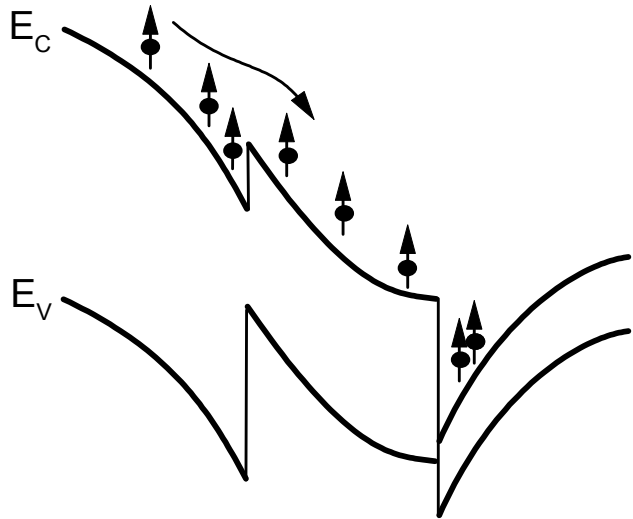
\end{document}